\documentclass[a4paper,11pt,twoside]{article}

\usepackage{amsmath}
\usepackage{amssymb}
\usepackage{epsfig}

\newcommand{\TeV}{\,\mathrm{TeV}}
\newcommand{\GeV}{\,\mathrm{GeV}}
\newcommand{\MeV}{\,\mathrm{MeV}}
\newcommand{\keV}{\,\mathrm{keV}}
\newcommand{\eV}{\,\mathrm{eV}}
\newcommand{\ns}{\,\mathrm{ns}}
\newcommand{\us}{\,\mathrm{\mu s}}
\newcommand{\fracwithdelims}[4]{\left#1 \frac{#3}{#4} \right#2}
\newcommand{\ord}[1]{\mathcal{O}\left( #1 \right)}

\newcommand{\capdef}{}
\newcommand{\mycaption}[2][\capdef]{\renewcommand{\capdef}{#2}%
        \caption[#1]{{\itshape #2}}} 
\makeatletter
\renewcommand{\fnum@table}{\textbf{\tablename~\thetable}}
\renewcommand{\fnum@figure}{\textbf{\figurename~\thefigure}}
\makeatother

\newlength{\myem}
\newcommand{\sep}[1]{#1}
\newcounter{mysubequation}[equation]
\renewcommand{\themysubequation}{\alph{mysubequation}}
\newcommand{\mytag}{\stepcounter{mysubequation}%
\tag{\theequation\protect\sep{\themysubequation}}}
\newcommand{\globallabel}[1]{\refstepcounter{equation}\label{#1}}

\makeatletter
\renewcommand{\section}{\@startsection{section}{1}{0em}%
	{-3.5ex \@plus -1ex \@minus -.2ex}%
	{2.3ex \@plus.2ex}%
	{\normalfont\large\bfseries}}
\renewcommand{\subsection}{\@startsection{subsection}{2}{0em}%
	{-3.25ex\@plus -1ex \@minus -.2ex}%
	{1.5ex \@plus .2ex}%
	{\normalfont\bfseries}}
\renewcommand{\subsubsection}%
	{\@startsection{subsubsection}{3}{0em}%
	{-3.25ex\@plus -1ex \@minus -.2ex}%
	{1.5ex \@plus .2ex}%
	{\normalfont\itshape}}
\makeatother

\newcommand{\decay}{\pi^+\rightarrow\mu^+X}
\newcommand{\dmg}{\delta m_X}
\newcommand{\dms}{\Delta m_{\text{KK}}}

\newcommand{\MP}{M_{\text{Pl}}}

\newcommand{\Z}{\mathbb{Z}}

\newcommand{\mbb}{{\hat{m}}}
\newcommand{\Fig}[1]{Fig.~\ref{fig:#1}}

\newcommand{\Eq}[1]{Eq.~(\ref{#1})}
\newcommand{\eq}[1]{eq.~(\ref{#1})}

\newcommand{\dash}{\,\text{--}\,}
\newlength{\phantomlength}
\newcommand{\phantomheigth}[1]{%
\settowidth{\phantomlength}{$\displaystyle #1$}%
\phantom{#1}\hspace*{-\phantomlength}}

\newcommand{\bea}{\begin{eqnarray}}
\newcommand{\eea}{\end{eqnarray}}

\def\a{\alpha}
\def\b{\beta}
\def\g{\gamma}

\def\d{\delta}
\def\e{\epsilon}

\def\f{\phi}

\def\m{\mu}
\def\n{\nu}

\def\p{\pi}

\def\r{\rho}
\def\s{\sigma}

\def\t{\tau}

\def\x{\xi}

\def\D{\Delta}

\def\G{\Gamma}

\def\pt{\partial}
\def\cL{\cal L}

\newcommand{\OX}{Department of Physics, Theoretical Physics,
University of Oxford, Oxford OX1\hspace{0.2em}3NP, UK}

\newcommand{\preprintdate}{April 2000}
\newcommand{\preprintnumber}{OUTP--00--15P}
\newcommand{\hepnumber}{hep-ph/0004130} 
\newcommand{\titletext}{A Brane--World Explanation of the KARMEN Anomaly} 
\newcommand{\authortext}{\large Andr\'e Lukas and Andrea Romanino
\medskip\\\em\normalsize 
\OX}
\newcommand{\abstracttext}{
Motivated by the anomaly in the KARMEN experiment, we study new
possibilities for brane--world models of neutrino masses. We show that
the KARMEN anomaly can be understood in the context of a
six--dimensional brane--world model. The fine--tuning problem
associated with the single--particle interpretation of the anomaly
is thereby alleviated. Our model shows some interesting properties
of general interest for brane--world neutrino physics. 
In particular, the size of the brane--bulk mixing decouples from the
scales relevant for atmospheric and solar neutrino oscillations.}

\title{
\normalsize
\begin{tabular}[t]{l}\hepnumber\\\preprintdate\end{tabular}
\hspace*{\fill}
\begin{tabular}[t]{l}\preprintnumber\end{tabular}
\vspace{3\baselineskip}\\\Large\bfseries\titletext\bigskip}
\author{\begin{minipage}[t]{0.8\textwidth}
\normalsize\centering\authortext
\end{minipage}}
\date{}

\begin{document}

\bigskip
\maketitle
\begin{abstract}\normalsize\noindent\abstracttext\end{abstract}
\normalsize\vspace{\baselineskip}

\section{Introduction}
\label{sec:introduction}
 
\noindent
Two recent developments, namely the discovery of brane--world
models~\cite{Horava:96a,Witten:96a,Horava:96b,Lukas:98b,%
Arkani-Hamed:98a,Antoniadis:98a,Kakushadze:98a} and the flexibility in
choosing fundamental scales and sizes of additional
dimensions~\cite{Witten:96a,Lykken:96a,Arkani-Hamed:98a,Antoniadis:98a},
have led to considerable activity exploring new options for particle
phenomenology. Both developments have a natural place in string-- or
M--theory which leads to interesting connections between the leading
candidate for a fundamental theory and particle physics.

Much of the interesting phenomenology of brane--world models is
associated with the Kaluza--Klein modes that originate from large,
gravity--only additional dimensions. Particularly the graviton and its
Kaluza--Klein excitations are of interest in this
context~\cite{Arkani-Hamed:98a}.  Neutrino physics is another branch
that is likely to be effected in brane--world models. This is because
higher--dimensional bulk fermions lead to Kaluza--Klein towers of
standard model singlets that may be interpreted as sterile
neutrinos~\cite{Dienes:98a,ADDM,Dvali:99a,BCS}. Most of
the work in brane--world neutrino physics to date has been focusing on
rather specific models. Concretely, most of the explicit studies have
been carried out for simple five--dimensional models with a special
form of the action that, for example, does not include bulk mass terms
(see, however, ref.~\cite{Dienes:98a}, where Majorana mass terms have
been considered in the context of orbifold models).

\medskip

Brane--world models with large additional dimensions have led to a
number of exciting predictions, such as the modification of gravity at
small distances~\cite{Arkani-Hamed:98a}, that may be testable in the
future.  In this paper, we would like to pursue the idea that a signal
originating from brane--world physics may have already been observed.
Specifically, we will show that the anomaly in the KARMEN neutrino
experiment~\cite{Armbruster:95a,KARMEN} can be explained in the
context of brane--world models. This will also lead us to explore new
possibilities for brane--world models of neutrino masses.  That is, we
will examine a number of new model building aspects in brane--world
theories of neutrino masses.

The KARMEN neutrino experiment studies neutral current interactions of
neutrinos that originate from the decay chain $\p^+\rightarrow \m^+$
at rest.  The KARMEN experiment has a unique time structure that
allows one to study the number of neutrino events in the detector as a
function of time after the production of the initial $\p^+$. It is
this distribution that shows an anomalous peak at a specific time
after $\p^+$ production. The KARMEN anomaly can be explained in terms
of a hypothetical $X$ particle~\cite{Armbruster:95a} produced in a
rare decay $\p^+\rightarrow\m^+X$. It has also been
shown~\cite{Barger:95a} that the $X$ particle can be interpreted as a
sterile neutrino. While postulating such a particle leads to a
satisfactory fit of the experimental data it has a theoretically
unappealing feature. To match the observed peak, the $X$ particle has
to have a very specific mass $m_X$ that is fine--tuned to the mass
difference $m_{\p}-m_\m$ with an accuracy of approximately
$10^{-4}$. That is, $m_X/(m_\pi-m_\mu)\simeq 1-1.8\,10^{-4}$. There is
no apparent theoretical explanation for such a coincidence of masses.

Here is where brane--world physics might come into play. Suppose, that
a brane--world model leads to a Kaluza--Klein tower of sterile
neutrinos. Clearly, if the spacing is sufficiently small,
it is relatively more probable for one of the particles in this
tower to fall into the narrow mass range relevant for the KARMEN
experiment than it is for a single particle. In the extreme case,
where the spacing of the tower is of the order of the mass gap
available for the $X$ particle, the fine--tuning problem is solved
completely.

\medskip

In this paper, we will show that a brane--world model explaining the
KARMEN anomaly does indeed exist. An analysis of the basic
requirements singles out two specific choices of dimensions and
scales. The first one corresponds to a five--dimensional model with
intermediate string scale and the scale of the fifth dimension in
the $\keV$ range. The second one corresponds to a six--dimensional
model with a $\TeV$ string scale
and the scale of the two additional dimensions in the $\MeV$ range.
We will focus on the second case and construct an explicit
six--dimensional model that meets all the requirements.
A major phenomenological constraint on this model is that the other
Kaluza--Klein sterile neutrinos do not distort the KARMEN spectrum
in a way inconsistent with the experiment. We will show that this can
be avoided by making the spacing of the Kaluza--Klein tower sufficiently
large. Unfortunately then, the fine--tuning problem cannot be solved
completely in the way indicated above but is still improved by two
orders of magnitude with respect to the single particle interpretation.
At the same time, the effect of the Kaluza--Klein modes lighter than the
$X$ particle, provide a way of how our proposal could be
experimentally distinguished from the single particle proposal. These
particles contribute to the KARMEN spectrum at short times and,
depending on the parameters of the model, the corresponding signal may
be detectable in future experiments.

\medskip

Our six--dimensional model also shows a number of interesting
properties that are of general importance for brane--world models of
neutrino masses. In the simplest version of the model with conserved
lepton number, for example, we find three massless eigenstates that
are generally nontrivial linear combinations of the electroweak
eigenstates and the bulk neutrinos. The admixture of bulk neutrinos
is controlled by a six--dimensional Dirac mass and can be made small
or large while keeping the states massless. In contrast, in the
models considered so far, massless modes were either not present
or decoupled from the ordinary neutrinos. Furthermore, in those
models, light neutrino masses and their mixing with bulk neutrinos are
usually controlled by the same parameters. Our model shows that it is
actually possible to decouple these two quantities. More generally,
it is illustrated that the inclusion of all Lorenz--invariant
(or even Lorenz non--invariant) mass terms in the higher--dimensional
theory opens up new possibilities for brane--world neutrino phenomenology.

\section{The KARMEN anomaly and its brane--world interpretation}
\label{sec:karmen}

\noindent
The KARMEN experiment at the Rutherford Appleton Laboratory studies
charged and neutral current interactions of neutrinos from the
$\pi^+\rightarrow\mu^+$ decay chain at rest. The unique feature of the
neutrino flux is its time structure. The primary pions are produced
in $0.5\us$ long pulses\footnote{More
precisely, the $0.5\us$ pulses are made of two shorter $100\ns$ long
pulses separated in time by $330\ns$.} with a frequency of
$50\,\text{Hz}$ by the ISIS synchrotron proton beam. Due to the short
pion lifetime, $\tau_\pi\simeq 26\ns$, the muon neutrinos from the
$\pi^+\rightarrow\mu^+\nu_\mu$ decay represent the prompt component of
the neutrino beam, which exhausts itself within a few $\tau_\pi$ after
the pion pulse. Hence, in the time window $(0.6$--$10.6)\us$
after proton beam on target the neutrino beam is exclusively composed
out of $\nu_e$ and $\bar\nu_\mu$ that originate from the slow decay
($\tau_\mu\simeq 2.2\us$) of the muons produced in the first
$0.5\us$. Consequently, the time spectrum of the number $n$ of
detector events in this time window is expected to be described by
\begin{equation}
\label{spectrum}
\frac{dn}{dt} = A e^{-t/\tau_\mu}+B\; .
\end{equation}
Here $A$ is a time--independent\footnote{The convolution of the muon
exponential decay rate with the time spectrum of muon production is
again an exponential with the same time constant.}  rate constant
depending, among other things, on the muon production rate, on the
target cross section and on the detector efficiency. The constant $B$
represents the event rate associated with a time--independent
background.

\begin{figure}
\begin{center}
\epsfig{file=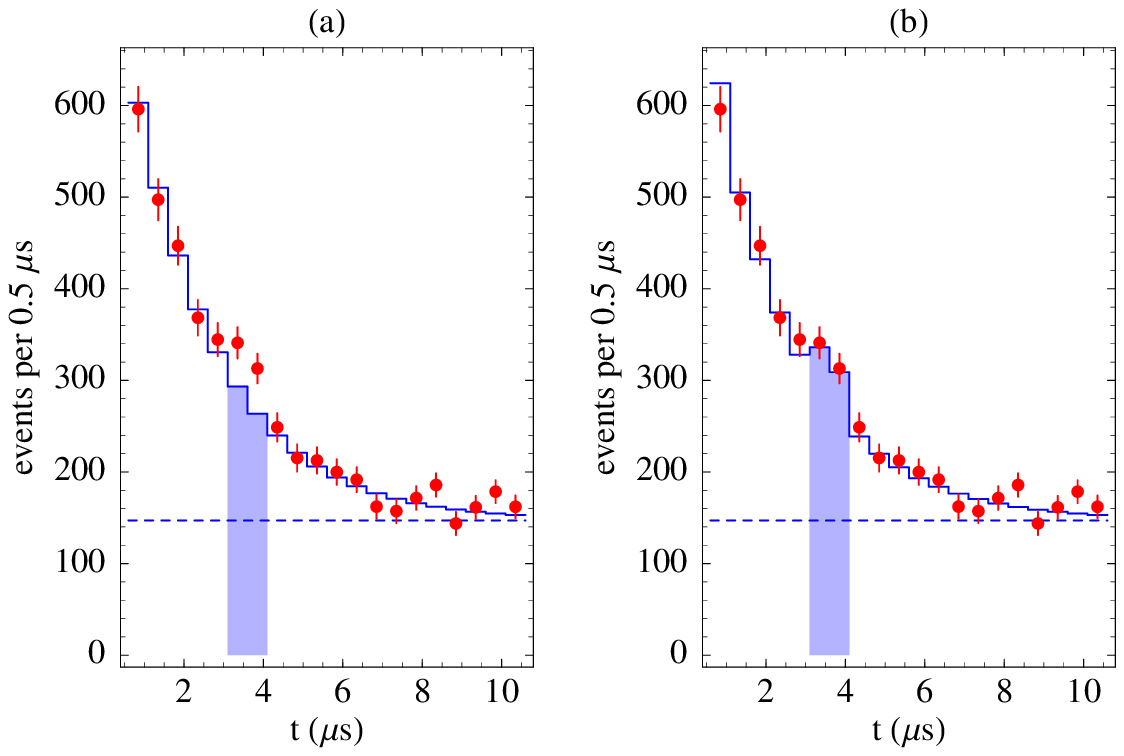,width=0.75\textwidth}
\end{center}
\mycaption{Time distribution of events in the time window $(0.6\dash
10.6)\us$ after beam on target and in the energy interval $(10\dash
36)\MeV$. The time interval is divided in 20 bins. The dots represent
the measured events. The dashed line indicates the measured
background. In Figure (a) the solid histogram shows the best fit of
all bins except the 6th and the 7th (shadowed bins) using the event
rate in \eq{spectrum}. In Figure (b) the signal associated with a
Kaluza--Klein tower of states corresponding to $1/R\sim 15\MeV$ is
added (see text).}
\label{fig:spe} 
\end{figure}

The KARMEN time--anomaly~\cite{Armbruster:95a,KARMEN} manifests itself
in a discrepancy between the expected time spectrum~(\ref{spectrum})
and the measured one. This discrepancy is mainly due to a peak at
approximately $3.6\us$ after beam on target. Such a peak is clearly
visible in \Fig{spe}a. In this plot the time window $(0.6\dash 10.6)\us$
has been divided into 20 bins and the number of events for each bin
(falling into the energy range from $10$ to $36\MeV$) has been
indicated by a dot. In order to estimate the significance of the
discrepancy and for later comparison with results based on our model
we have performed a simple analysis of the data shown in
\Fig{spe}a. Specifically, we have carried out a $\chi^2$ fit of the
data set in \Fig{spe} assuming an event rate as in
\eq{spectrum}. Background measurements on a wider time interval fix
the parameter $B$ as indicated by the dashed line in \Fig{spe}. The
best fit results in $\chi^2_{\text{min}}= 29$ for 19 degrees of
freedom, corresponding approximately to a $2\sigma$
discrepancy. However, the significance of the discrepancy is higher if
one considers only the two bins around $t=3.6\us$. This is appropriate
because an explanation of the anomaly around $t=3.6\us$ by statistical
fluctuation is strongly disfavored.  The peak around $t=3.6\us$,
already clearly present in the 1995 data~\cite{Armbruster:95a}, has,
in fact, been confirmed by the subsequent KARMEN2
data~\cite{KARMEN}. Once the two bins around $t=3.6\us$ have been
excluded, the remaining data is very well described by the
distribution in \eq{spectrum}. The best fit leads to
$\chi^2_{\text{min}}= 14$ with 17 degrees of freedom,
corresponding to the solid line in \Fig{spe}a. According to such a
fit, a total of 557 events are expected in the two
unfitted bins. As shown in~\cite{KARMEN}, the measured 659 events
represent a deviation at the level of approximately $4\sigma$. Since
the time--anomaly appears to be statistically significant and any
attempt to explain its peak structure by systematic effects failed so
far, it is worth to speculate about its physical origin.

A simple description of the time spectrum can be achieved by assuming
that the peak around $t=3.6\us$ is associated with the decay of a
slowly moving massive neutral particle ``$X$'' in the detector. Such a
particle could, for example, be produced in the rare decay
$\pi^+\rightarrow\mu^+X$~\cite{Armbruster:95a}. The position of the
peak then measures the time of flight of the particle and hence its
velocity. According to the recent KARMEN fits~\cite{KARMEN} the
velocity is given by $\beta_X\simeq 0.0162$.  This velocity, in turn,
determines the time the particle spends in the detector and, together
with the time structure of the proton beam, the width of the
signal. This width is found to be in nice agreement with the observed
width of the peak. More generally, the slow particle interpretation is
supported by a detailed analysis of the whole data set. Due to the
very good time and spatial resolution of the detector, $\Delta
t<2\ns$, $\Delta x$, $\Delta y$, $\Delta z< 15\,\text{cm}$, an
analysis of the correlation between time and position of the events is
possible. Such an analysis is in agreement with the slow particle
hypothesis and provides the most precise measurement of the $X$
particle velocity leading to the value given above~\cite{KARMEN}.

Once the signal is attributed to the decay of a slow particle, a rare
$\pi^+\rightarrow\mu^+X$ decay represents the simplest possibility for
the production of the particle. Since the time width of the anomaly is
smaller than the muon lifetime, it is natural to assume that its
production is not associated to muon decay (see
however~\cite{Gninenko:98a}). Moreover, the production in the $\pi^+$
decay allows to account for the slow, fixed velocity in terms of a
small available phase space (unlike, for example, the production by
the proton interactions with the target). Finally, the energy
deposited by the particle in the detector is found to be less than
about $35\MeV$~\cite{KARMEN}. This is what is expected from the decay
of a particle produced in the decay $\pi^+\rightarrow\mu^+X$, whose
mass $m_X$ has to be less than $m_\pi-m_\mu\simeq 33.9\MeV$. On the
other hand, if the production process were $\pi^+\rightarrow e^+X$, the
slow $X$ particle would have to have a mass $m_X\simeq 137\MeV$ and
would release more energy than measured (unless having a muon or
another massive particle in the final state is for some reason
favored). As for the decay, the simplest possibilities for the
visible decay of a particle produced in the $\pi^+\rightarrow\mu^+X$
channel are $X\rightarrow e^+e^-\nu$ and $X\rightarrow\gamma\nu$,
where $\nu$ generically represents a neutral light particle. A visible
decay mainly through $X\rightarrow\gamma\nu$ thereby producing
mono--energetic photons is highly
disfavored~\cite{KARMEN}. Consequently, in this paper we will
consider the possibility that the $X$ particle is produced in the
decay $\pi^+\rightarrow\mu^+X$ decay and the anomaly is mainly due to
its visible decay $X\rightarrow e^+e^-\nu$. If the $X$ particle
lifetime is larger than $3.6\us$, as it will be in our case, the
KARMEN data requires that

\begin{equation}
\label{constr}
\text{BR}(\decay)\cdot\Gamma_{\text{vis}} =
(1\dash 2)\,10^{-11}\,\text{s}^{-1},
\end{equation}
where $\text{BR}(\decay)$ is the branching ratio of the decay $\decay$
and $\Gamma_{\text{vis}}$ is the partial width of the visible $X$
decays. 
\medskip

The first proposal for the nature of the hypothetical $X$ particle was
made in~\cite{Barger:95a}. There, it was identified with a sterile
neutrino that mixes with the Standard Model (SM) neutrinos. The same
possibility was considered in~\cite{P,Govaerts:96a}. Supersymmetric
scenarios were studied in~\cite{Choudhury:96a,Choudhury:99a}.  From a
theoretical point of view, light sterile neutrinos are welcome if
their lightness can be accounted for. Whereas the mass of the SM
neutrinos is protected by the electroweak symmetry (and the absence of
fundamental isovector Higgses), the SM symmetries alone cannot explain
why sterile neutrinos should be lighter than, say, the Planck mass. It
is therefore useful to consider sterile neutrinos in the context of a
specific framework. In this paper, we will be dealing with sterile
neutrinos that arise in brane--world models with large additional
dimensions. Concretely, we consider the possibility that the KARMEN
anomaly is due to a sterile neutrino that is part of a tower of
Kaluza--Klein excitations associated with a SM--singlet fermion
propagating in $4+d$ dimensions. A possible explanation for small
sterile neutrino masses is available if these models are viewed in the
context of string theory.  Then, bulk fermions arise naturally as
superpartners of moduli fields. Perturbatively, moduli are flat
directions of the theory, and, hence, their fermionic partners are
massless before supersymmetry breaking. After including supersymmetry
breaking effects those fermions will receive a mass related to the
supersymmetry breaking scale. Depending on the details, this might
well provide a reason for small bulk fermion masses. The next sections
will be devoted to a detailed study of such models, particularly in
six dimensions. In the remainder of this section, we will describe some
general features of our six--dimensional model from a phenomenological
viewpoint and summarize its implications for the KARMEN experiment.

\medskip

Let us consider six--dimensional Dirac fermions as the origin of the
sterile neutrinos. For simplicity we will consider only one such Dirac
fermion here, although our later treatment will be more general.  From
a four--dimensional perspective, this field can be described by a
Kaluza--Klein tower of sterile Dirac neutrinos $\nu_n$ labeled by two
integer numbers $n=(n_1,n_2)$. Assuming, for simplicity, that the two
extra dimensions have equal size $2\pi R$, the mass spectrum of the
Kaluza--Klein tower is given by by
\begin{equation}
\label{masses}
m_n = \sqrt{\mu^2+\frac{n^2}{R^2}},
\end{equation}
where $n^2 = n_1^2+n_2^2$ and the mass parameter $\mu$ originates from
a six--dimensional Dirac mass term. Of course, to have a candidate for
the $X$ particle among these Kaluza--Klein fermions we have to require
that the mass parameter $\m$ is smaller than mass $m_X$ of the $X$
particle. This requirement precisely corresponds to the problem of
keeping sterile neutrinos light. With the above string theory
explanation for this in mind, we assume that $\m < m_X$. Then, we may
have a specific sterile neutrino $X=\nu_{n_X}$ in the Kaluza--Klein
tower that we would like to associate with the KARMEN anomaly.  From
\eq{masses} we can compute the density of states $\rho(m_X)$ at the
scale $m_X$, defined as the average number of states $dn$ per mass
interval $dm$. Neglecting the mass $\m$ for an order of magnitude
estimate, this density can be computed from eq.~\eqref{masses} to
be~\footnote{\label{foot:degeneracy}In our simplified model the
$n$--th state is eight times degenerate if $n_1\neq0$, $n_2\neq 0$,
$n_1\neq n_2$. In a general situation where the size of the two extra
dimensions are different and all possible six--dimensional mass terms
are included, the degeneracy is lifted.}
\begin{equation}
\label{density}
\rho(m_{\rm X}) \simeq 2\pi R^2 m_{\rm X}\; .
\end{equation}
In other words, the average separation between two subsequent states at
$m_{\rm X}$ in a mass--ordered list is given by
\begin{equation}
\label{separation}
\dms = \frac{1}{\r (m_{\rm X})} \simeq \frac{1}{2\pi R^2 m_{\rm X}}.
\end{equation}
The last equation highlights an interesting feature of our framework.
Besides being able to account for the existence of light sterile
neutrinos, it allows us to alleviate the fine--tuning problem
associated with the slow particle interpretation of the KARMEN
anomaly. The fine--tuning problem arises because the slowness of the
$X$ particle requires its mass to be only slightly lower than the
difference $m_\pi-m_\mu\simeq 33.9\MeV$. More precisely, in order to
be produced with a velocity $\beta_X\simeq 0.0162$, the $X$ particle
mass $m_X$ must be
\globallabel{FT}
\begin{equation}
m_X = m_\pi-m_\mu-\dmg,
\end{equation}
where
\begin{equation}
\dmg \simeq \beta^2_X\,(m_\pi-m_\mu)\frac{m_\pi}{2 m_\mu}\simeq 6\keV.
\end{equation}
This means that $m_X$ is fine tuned to the $m_\pi-m_\mu$ mass
difference with a precision of $1.8\times 10^{-4}$! On the other hand the
fine tuning problem is significantly alleviated in the approach we
propose, since the $X$ particle is just one of the many excitation of the
six--dimensional fermion. It is therefore relatively more probable
that one out of the many states has a mass that falls into the
critical range.

In principle, this framework allows us to completely eliminate the fine
tuning problem. In fact, the problem disappears if the
average separation between subsequent states is about twice the
mass range available for the $X$ particle, that is,
$\dmg = m_\pi - m_\mu - m_X \sim \dms/2$. From eq.~\eqref{separation} this
happens if
\begin{equation}
\label{Rnat}
\frac{1}{R} =(2\pi m_X \dms)^{1/2} \sim (4 \pi m_X \dmg)^{1/2} \simeq
1.6\MeV.
\end{equation}
In this case, the production of the particles heavier than $X$ in the
$\pi^+\rightarrow\mu^+\nu_n$ decay would be kinematically
forbidden\footnote{This assumes that there is no local inhomogeneity
in the density of states, so that all states heavier than $X$ are
indeed heavier by a margin larger than the mass gap $\dmg$ of the X
particle.}. However, the states lighter than $m_{\rm X}$, being faster
than the $X$ particle, modify the time spectrum in the window
$(0.6\dash 3.6)\us$. In order compute the effect of those lighter
states on the time spectrum, one has to take into account that they
are produced more copiously due to the larger phase space available.
However, their signal is shorter because they spend less time inside
the detector.  A simulation taking into account the detailed
time--shape of the signal associated with the lighter particles shows
that the KARMEN data is in very good agreement with the mass spectrum
associated with an inverse radius of about $15\MeV$, namely about 10
time larger than the natural value in~(\ref{Rnat})\footnote{The
simulation assumes that the degeneracy mentioned in
footnote~\ref{foot:degeneracy} is completely lifted.}. \Fig{spe}b
shows the corresponding best fit (solid line). Once the mass spectrum
has been fixed by the choice of $1/R$, the overall normalization of
the sterile neutrinos signal and the constant $A$ in \eq{spectrum} are
fitted to the data. The best fit shown in Figure has a
$\chi^2_{\text{min}} = 15.8$ for 18 degrees of freedom. The fine
tuning problem is alleviated. The maximum density of states at the
$m_X$ scale is about 40 times larger than $(m_\pi-m_\mu)^{-1}$, making
this framework about 40 times more natural than the standard framework
with a single particle. However, a significant fine tuning is still
needed.

The discussion above points out a peculiar feature of our
proposal. The $X$ particle comes with a number of particles lighter
than $m_X$ whose number depends on the radius of the additional
dimensions. For large values of the inverse radius, the lighter states
are separated by a large mass gap. In this case our model is
indistinguishable from models with a single sterile neutrino as far as
the KARMEN experiment is concerned. However, from the theoretical
point of view, it still provides a well motivated scenario for the
existence of a sterile neutrino with a mass of $33.9\MeV$ which, at
the same time, can account for the standard oscillation phenomenology
(see Section~\ref{sec:model}). On the other hand, if the average mass
gap $\dms$ between the states is smaller than about $5\MeV$, the
particles lighter than $X$ could give a small contribution to the
early KARMEN time spectrum. For example, with the parameters used for
the fit shown in \Fig{spe}b, corresponding to an average separation of
about $1\MeV$ between states, one gets a small contribution to the
number of events in the first time bin. Therefore, a detailed study of
the time distribution at early times after the end of the
proton pulse could test the model if the mass separation $\dms$
happens not to be too large. An upgraded detector with tracking
capability devoted to the study of the anomaly could certainly explore
this possibility.  Needless to say, the detection of a signal at early
times would represent a strong hint for the existence of a
brane--world.

\section{Scales and couplings}
\label{sec:scales}

In this section, we would like to analyze the scales and couplings of
the brane--world theory that are required for a solution of the KARMEN
anomaly along the lines explained above. This analysis
will provide us with an overview over the various models that may be
suitable for our purpose. In the following section, we will pick one
of the models found in this way and develop it in detail.

We start with a ten--dimensional string theory with string scale
$M_{\rm s}$. The six internal dimensions are schematically divided
into two groups. The first group of $d$ dimensions are the ones that
give rise to Kaluza--Klein neutrinos potentially relevant for the
KARMEN experiment. The other $6-d$ dimensions are those without any
direct relevance for the experiment. For example, a particular
dimension is not relevant in this sense if its inverse size is
much larger than the mass of the $X$ particle. For simplicity we
assume that the first $d$ dimensions have the same size characterized
by the radius $R$. This is likewise assumed for the remaining
$6-d$ dimensions where the corresponding radius is called $\r$. Given this
setup, there are three basic requirements to be satisfied.
First, the four--dimensional Planck constant $\MP$ specified by
\begin{equation}
 \left(\frac{\MP}{M_{\rm s}}\right)^2 = 16\p (2\p RM_{\rm s})^d
      (2\p\r M_{\rm s})^{6-d}
\end{equation}
should have the correct value. Except for this general condition we
have two more constraints that are specific to our problem.  
The first one is related to the probability of having a
Kaluza--Klein particle with mass in the relevant range for the
slow--particle interpretation of the KARMEN experiment. We would like
to significantly reduce the amount of fine tuning required. This means
that the parameter $\D$ defined as
\begin{equation}
 \frac{1}{\D} = \r (m_{\rm X})\dms\; 
\end{equation}
should be significantly smaller than $\ord{10^4}$, corresponding to
the fine--tuning of the single--particle solution.  We remind the
reader that $\r (m_{\rm X})$ is the density of Kaluza--Klein states at
$m_X=33.9\MeV$ and $\dms\simeq 6\keV$ is the mass gap between the $X$
particle mass and the $\pi$-$\mu$ mass difference. Note that, from the
distinction of dimensions made above, only the Kaluza--Klein modes
associated to $d$ dimensions are relevant for this density. For $N$
Dirac spinors in $4+d$ dimensions with a Dirac mass $\m$ it is given
by
\begin{equation}
 \r (m_{\rm X}) = \frac{d\,2^{[d/2]}\p^{d/2}}{\G (d/2+1)}NR^dm_{\rm X}
                  (m_{\rm X}^2-\m^2)^{d/2-1}\; .\label{density1}
\end{equation}
The final requirement originates from the constraint~\eqref{constr} on
the branching ratio and the visible decay width of the $X$ particle.
This constraint implies that there should be a non--negligible
mixing between the $X$ particle and the left--handed neutrinos.
In the models under consideration such a mixing is generated by a
mass term between brane and bulk fields that resides on the brane.
On dimensional grounds the associated mass parameter $\mbb$ can be written as
\begin{equation}
 \mbb = \frac{hv}{(2\p RM_{\rm s})^{d/2}}=\frac{hm_t}
     {h_t(2\p RM_{\rm s})^{d/2}}\; ,
\end{equation}
where $h$ is a brane--bulk Yukawa coupling, $v$ is the the standard
model Higgs VEV and $h_t$ and $m_t$ are the top Yukawa--coupling and
mass. Then, the mixing matrix element $U_{\n X}$ between the left--handed
neutrinos and the $X$ particle specified by
\begin{equation}
 U_{\n X} = \frac{\mbb}{m_X}
\end{equation}
should not be too small. Typically, from eq.~\eqref{constr} one has
$|U_{\n X}|\simeq 10^{-3}$. 

Altogether, we now have three constraints for our three fundamental
parameters $M_{\rm s}$, $R$ and $\r$. All other quantities appearing
in those constraint are either fixed or in a well--defined range. We
can therefore simply solve for $M_{\rm s}$, $R$ and $\r$ as a function
of the number of dimensions $d$. The fact that we can basically
determine all fundamental scales is quite remarkable and shows how
constrained the problem is. A priori, it is not clear at all that a
solution with sensible values of the scales exists. However, as
Table~\ref{tab1} shows, this is indeed the case. The quantity $y$
appearing in the table is defined as
\begin{equation}
 y = \frac{h}{10^3|U_{\nu X}|h_t}\; .
\end{equation}
\begin{table}
 \begin{center}
 \begin{tabular}{|c|c|c|c|}
 \hline
 &$(N\D y^2)^{-1/d}\times$&$(N\D )^{-1/d}\times$&
 $(N\D )^\frac{d-8}{d(6-d)}y^\frac{-16}{d(6-d)}\times$\\
 &$M_{\rm s}$&$R^{-1}$&$\r^{-1}$\\\hline
 $d=1$&$4.5\times 10^7\GeV$&$12\keV$&$7.9\times 10^6\GeV$\\\hline
 $d=2$&$1.2 \TeV$&$1.6\MeV$&$461\MeV$\\\hline
 $d=3$&$23\GeV$&$5.6\MeV$&$28\keV$\\\hline
 \end{tabular}
 \end{center}
\caption{\em Shown are scales and dimensions relevant for an explanation of the
         KARMEN experiment. We present typical values of the string
         scale $M_{\rm s}$, the inverse radius $R^{-1}$ of the $d$
         additional dimensions associated with relevant
         Kaluza--Klein neutrinos and the inverse radius $\r^{-1}$ of
         the remaining $6-d$ dimensional internal space.}
 \label{tab1}
\end{table}
Let us discuss the various cases shown in table~\ref{tab1}. For one
additional dimensions $d=1$, we have a model with intermediate string
scale, one large additional dimension in the $\keV$ range and the
size of the other five dimensions close to the string scale. For two
additional dimensions, $d=2$, the situation is quite different. We
have a model with the string scale in the $\TeV$ range, two large
additional dimensions in the $\MeV$ range and the scale of the other
four internal dimensions being about two to three orders of
magnitude larger. In the case
of three additional dimensions, $d=3$, the typical string scale is already
quite low and one needs a large factor $N\D y^2$ to elevate it above
the required $\TeV$ limit. Therefore, this case only represents a
marginal possibility\footnote{Moreover, unlike for the other two
cases, the scale $\r^{-1}$ is smaller than the mass $m_{\rm X}$ of
the $X$ particle. Therefore, if the Kaluza--Klein neutrinos propagated
in those dimensions they will be relevant for the experiment,
contrary to our initial definition of those dimensions. Therefore, one
has to assume that this is not the case. Such an assumption is
probably not very appealing from the perspective of model--building.}.
Finally, one can show that models with more than three additional
dimensions, $d>3$, lead to a string scale below $\TeV$ and are,
therefore, not viable in our context.

In summary, we have isolated two cases which may lead to an
explanation of the KARMEN anomaly in the context of a brane--world
model. The first case represents an effective five--dimensional
model with intermediate string scale and the remaining five dimensions
being close to the string scale. The second one is a model with a $\TeV$
string scale, two large dimensions in the $\MeV^{-1}$ range and the
scale of the remaining four dimensions being two to three orders of
magnitude larger. Therefore, this model is effectively
six--dimensional in an intermediate energy range. We stress that the
existence of these two possibilities, particularly the one with a
$\TeV$ string scale, was by no means guaranteed and is due to a
rather fortunate conspiracy of the various mass scales in the problem.
While explicit examples can be constructed in both cases, in this
paper we will focus on the six--dimensional case with a $\TeV$ string scale. 

\section{An explicit six--dimensional example}
\label{sec:model}

In this section, we would like to present an explicit brane--world
model that realizes the ideas outlined above.

It is clear from the above discussion of scales that it is sufficient
for our purpose to construct an effective six--dimensional brane--world
model, valid for energies below $\r^{-1}$. The Kaluza--Klein modes
associated to the remaining four internal dimensions are too heavy to
be relevant within our context. More specifically, we would like to
couple a six--dimensional bulk with right--handed neutrinos
and a three--brane that carries the standard--model fields.
Such a model should then be analyzed in detail with respect to its
consequences for the KARMEN anomaly. Furthermore, as we will see, the
model illustrates some general features that arise in brane--world
models of neutrino masses which have not been considered so far.

The six--dimensional bulk action of the model is specified by
\begin{equation}
 S_{\rm bulk} = \int d^4xd^2y\left[ \bar{\Psi}_I\G^A i\pt_A\Psi_I-
                (\m_{IJ}\bar\Psi_{LI}\Psi_{RJ}+\mbox{h.c.})\right]\;
                .
 \label{Sbulk}
\end{equation}
Focusing on the relevant Yukawa couplings between bulk and brane
fields, the brane action reads
\begin{equation}
S_{\rm brane} = \int_{\{y=0\}}d^4x\left[ -\frac{h_{aIi}}{M_{\rm s}}
                \bar{\Psi}_{aI}L_iH +\mbox{h.c.}\right]\; .
\label{Sbrane} 
\end{equation}
We use coordinates $x^A$ with indices $A,B,\cdots = 0,\cdots ,5$ for
the total six--dimensional space--time. Four--dimensional coordinates $x^\m$
are indexed by $\m ,\n ,\cdots = 0,1,2,3$. The remaining two coordinates
are denoted by $y^\a$, where $\a ,\b , \cdots =1,2$.
The signature of our metric is ``mostly minus''. We have introduced
$N$ six--dimensional Dirac fermions $\Psi_I$, where
$I,J,\cdots = 1,\cdots ,N$, in the bulk. Their left-- and right--handed
components are defined in the usual way by $\Psi_{R/L,I}=(1\pm
\G_7)\Psi_I$ where $\G_A$ are the six--dimensional gamma matrices and
$\G_7=\G_0\cdots\G_5$. The $4+2$ decomposition of these gamma matrices
can be written in the form
$\G_I=\{\g_\m\otimes{\bf 1}_2,i\g_5\otimes\s_\a\}$. Here $\g_\m$ are
the four--dimensional Dirac matrices and $\g_5=-i\g_0\g_1\g_2\g_3$. The
two--dimensional Dirac--matrices $\s_\a$ can be identified with the
first two Pauli matrices. Defining the two--dimensional chirality
operator by $\s_3\equiv -i \s_1\s_2$ we have the relation
$\G_7=\g_5\otimes\s_3$ between six--, four-- and two--dimensional
chiralities. For later purposes we also define six--dimensional charge
conjugation by $\Psi_I^c=C^{-1}\bar{\Psi}_I^T$, where the charge
conjugation matrix $C$ is specified by the relations
$(\G_A)^T=-C\G_AC^{-1}$, as is usual in even dimensions. Furthermore,
we have introduced Dirac mass terms with associated mass matrix
$\mu_{IJ}$ in the bulk. By a suitable redefinition of the bulk
fermions we can diagonalize this mass matrix. In the following, we
will use this diagonalized form
\begin{equation}
 \mu_{IJ}=\mbox{diag}(\mu_1,\cdots ,\mu_N)
\end{equation}
with real, positive eigenvalues $\mu_I$. We also choose the flavor
basis for the lepton doublets $L_i$ such that the charged lepton
Yukawa couplings are diagonal. In order to be able to couple the bulk
spinors to brane field we decompose each six--dimensional field
$\Psi_I$ into two four--dimensional Dirac spinors $\Psi_{aI}$ where
$a=+,-$ indicates the internal two--dimensional chirality of the
component. The three--brane is taken to be located at $y=0$ in the
internal space and carries, among the other standard model fields, the
lepton doublets $L_i$, where $i,j,\cdots = e,\mu,\tau$, and the Higgs
doublet $H$. These fields have a Yukawa coupling to the two components
$\Psi_{aI}$ of the bulk spinors introduced above where the
dimensionless coupling constants are denoted by $h_{aIi}$. Note that,
from the dimensionality of the bulk spinors, the corresponding
operator are suppressed by one power of the string scale $M_{\rm s}$.

In summary, our model consists of $N$ ``right--handed neutrino'' bulk
spinors with a Dirac mass in six dimensions and Yukawa couplings to
the lepton doublets located on the three--brane. Our model, as stands,
has a lepton number $U(1)$ symmetry with $L_i$ and $\Psi_I$ each
carrying one unit of charge. We would like to impose this $U(1)$
symmetry (or an appropriate discrete subgroup thereof) on our model.
This forbids the other mass terms and couplings that we could have
written in our action. In particular, it forbids the bulk Majorana
mass terms $\bar{\Psi^c}_I\Psi_J$ and $\bar{\Psi^c}_I\G_7\Psi_J$ which
would be allowed by six--dimensional Lorenz invariance.  Furthermore,
it forbids the brane--bulk coupling $\bar{\Psi^c}_{aI}L_iH$ that would
be allowed from four--dimensional Lorenz invariance. However, we
stress that, at this stage, nothing forbids the bulk Dirac mass
term. It has, therefore, been included in the above action.

\medskip

Next, we would like to work out the four--dimensional effective action
of our model. We compactify the additional dimensions on a two--dimensional
torus that, for simplicity, is taken to be rectangular and with equal
radii $R$ in both direction. With the Higgs vacuum expectation value
$v$, we introduce the mass parameters
\begin{equation}
 \mbb_{aIi} = \frac{h_{aIi}v}{2\p RM_{\rm s}}\; .
\end{equation}
We expand the bulk spinors into Kaluza--Klein modes as
\begin{equation}
 \Psi (x,y) = \frac{1}{2\p R}\sum_{n\in\Z^2}\Psi_{In}(x)
              \exp\left(\frac{in_\a y^\a}{R}\right)\label{exp}
\end{equation}
where $n=(n_1,n_2)$. As we did with with the spinor $\Psi_I$ before,
we decompose each of its Kaluza--Klein modes $\Psi_{In}$ into two
four--dimensional Dirac spinors $\Psi_{aIn}$, where $a=+,-$.
Each of these four--dimensional Dirac spinors is decomposed further
into Weyl spinors according to
\begin{equation}
 \Psi_{aIn} =\left(\begin{array}{c}\x_{aIn}^c\\\eta_{aIn}
             \end{array}\right)\; .
\end{equation}
The charge conjugated Weyl spinor $\x^c$ is defined as
$\x^c=\e\x^*$ where $\e$ is the two--dimensional epsilon--symbol.
As a result, for each bulk spinor $\Psi_I$ and each fixed mode number
$n$ we obtain two four--dimensional Dirac spinors $\Psi_{aIn}$ or,
equivalently, four left--handed Weyl spinors $\x_{aIn}$, $\eta_{aIn}$,
where $a=+,-$. With these definitions, the mass terms in the four--dimensional
effective Lagrangian read
\bea
 \cL_{\rm bulk} &=& \sum_{n\in \Z^2}\left[m_n\x_{+In}\eta_{-In}
                -m^*_n\x_{-In}\eta_{+In}+\mu_I(\x_{+In}\eta_{+In}+
                \x_{-In}\eta_{-In})+\mbox{h.c.}\right] \label{Lbu}\\
 \cL_{\rm brane} &=& \sum_{n\in \Z^2}\left[m_{+Ii}\x_{+In}+
                   m_{-Ii}\x_{-In}\right]\n_i+\mbox{h.c.}\; .\label{Lbr}
\eea
Here, the Kaluza--Klein masses $m_n$ are defined by
\begin{equation}
 m_n = \frac{in_1+n_2}{R}\; .
\end{equation}
To interpret the structure of masses given above it is useful to
diagonalize the bulk Lagrangian. Since for each mode number $n$ and
bulk flavor $I$ we have two degenerate Dirac spinors, there is a
degree of arbitrariness in the choice of the eigenstates. To keep the
brane Lagrangian unchanged, we rotate the spinors $\eta$ only. We
therefore introduce the linear combinations
\begin{equation} 
{\eta '}_{+In} = \frac{\mu_I\eta_{+In}+m_n\eta_{-In}}
{\sqrt{\mu_I^2+|m_n|^2}}\; ,\qquad {\eta '}_{-In} =
\frac{\mu_I\eta_{-In}-m_n^*\eta_{+In}} {\sqrt{\mu_I^2+|m_n|^2}}
\end{equation}
which allows us to write the bulk Lagrangian in the form
\begin{equation}
 {\cal L}_{\rm bulk} = \sum_{n\in \Z^2}\sqrt{\m_I^2+|m_n|^2}
                \left[\x_{+In}{\eta '}_{+In}+
                \x_{-In}{\eta '}_{-In}+\mbox{h.c.}\right]\; .
\end{equation}
From this we read off the following mass matrix
\begin{equation}
 {\cal M} = 
\begin{matrix}
 \nu\\\vdots\\ 
\phantomheigth{\sqrt{\displaystyle\m^2+|m_n|^2}} \x_{+n} \\ 
\phantomheigth{\sqrt{\displaystyle\m^2+|m_n|^2}} {\eta'}_{+n} \\ 
\phantomheigth{\sqrt{\displaystyle\m^2+|m_n|^2}} \x_{-n} \\
\phantomheigth{\sqrt{\displaystyle\m^2+|m_n|^2}} {\eta '}_{-n} \\ 
\vdots
\end{matrix}
\begin{pmatrix}
  0&\cdots&\mbb_+^T&0&\mbb_-^T&0&\cdots\\
  \vdots&\ddots&&&&&\\
  \mbb_+&&0&\sqrt{\displaystyle\m^2+|m_n|^2}&0&0&\\
  0&&\sqrt{\displaystyle\m^2+|m_n|^2}&0&0&0&\\
  \mbb_-&&0&0&0&\sqrt{\displaystyle\m^2+|m_n|^2}&\\
  0&&0&0&\sqrt{\displaystyle\m^2+|m_n|^2}&0&\\
  \vdots&&&&&&\ddots
\end{pmatrix}\; ,
 \label{M}
\end{equation}
where for ease of notation we have used $\nu \equiv
(\nu_e,\nu_\mu,\nu_\tau)^T$, $\x_{an} \equiv
(\x_{a1n}\ldots\x_{aNn})^T$, $\eta'_{an} \equiv
(\eta'_{a1n}\ldots\eta'_{aNn})^T$ and $\mu=\text{diag}(\mu_1\ldots\mu_N)$.
Given this matrix, it is easy
to show that we have three exactly massless Weyl fermions $\n_{0i}$
given by
\begin{equation}
 \n_{0i} = (N^{-1})_{ij}\left[\n_j-\sum_{n,I,a}\frac{\mbb^*_{aIj}}
           {\sqrt{\m_I^2+|m_n|^2}}{\eta '}_{aIn}\right]\; .\label{massless}
\end{equation} 
Since in our context all the mass parameters are several orders of
magnitude above the eV scale, it is natural to attribute the solar and
atmospheric neutrino phenomenology to these massless eigenstates. The
occurrence of massless eigenstates is due to the U(1) symmetry we have
imposed. We will discuss below how a small breaking of this symmetry
can generate the small masses necessary to account for the standard
phenomenology of neutrino oscillations. As long as the states
$\n_{0i}$ are massless, the matrix $N$ in \eq{massless} can be
determined only up to a unitary transformation among the massless
eigenstates. However, $N$ will be fixed (up to phase rotations) once
small masses for the $\n_{0i}$ have been generated.  After integrating
over the Kaluza--Klein modes up to the cut--off $M_{\rm s}$ one finds
\begin{equation}
 (N^\dagger N)_{ij} \simeq \d_{ij}+\p\sum_{a,I}\mbb^*_{aIi}\mbb_{aIj}
                    \left[\frac{1}{\m^2}+R^2\ln\frac{M_{\rm
                    s}^2}{\m^2+1/R^2}\right]\; .
 \label{N}
\end{equation}
We are interested in a situation where the mass matrix has a
hierarchical structure satisfying $|\mbb_{aIi}|/\sqrt{\m_I^2+|m_n|^2}\ll 1$
for all states. In this case $(N^\dagger N)_{ij}\simeq\d_{ij}$ and
$N$ is approximately given by a unitary matrix $U$, that is  
\begin{equation}
 N\simeq U\; .
\end{equation}
An exact diagonalization of the mass matrix ${\cal M}$ shows that the
massive eigenstates can be organized in two Dirac spinors for each
bulk flavor index $I$ and each mode number $n$.  Assuming a
hierarchical structure of the mass matrix also simplifies the process
of finding such eigenstates since it implies that the mass
matrix~\eqref{M} can be diagonalized perturbatively. While this
reproduces eq.~\eqref{massless} for the massless states, it tells us
that for each bulk flavor index $I$ and each mode number $n$ the two
Dirac spinor eigenmodes $(\tilde\nu^c_{aIn},\nu_{aIn})^T$ are given
approximately by 
\globallabel{heavy}
\begin{align}
\tilde\nu^c_{aIn} &\simeq \xi_{aIn} \mytag \\
\nu_{aIn} &\simeq
\eta'_{aIn}+\sum_{i}\frac{\mbb_{aIi}}{\sqrt{\mu^2_I+|m_n|^2}} \nu_i\: ,
\mytag 
\end{align}
where $a=+,-$.
This equation holds up to corrections of order
$|\mbb_{aIi}|^2/(\m_I^2+|m_n|^2)$. Furthermore, in the same
approximation, these two spinors have a degenerate Dirac mass
\begin{equation}
 M_{In} = \sqrt{\m_I^2+|m_n|^2}\; .\label{kkmass}
\end{equation}
Given this information we can invert the formula~\eqref{massless}
and express the weak eigenstates $\n_i$ in terms of mass eigenstates
$\n_{0i}$, $\n_{aIn}$ as
\begin{equation}
\label{flavor}
 \n_i \simeq U_{ij}\n_{0j}+\sum_{n,I,a}\frac{\mbb^*_{aIi}}
           {\sqrt{\m_I^2+|m_n|^2}}\n_{aIn}
\end{equation}
up to small terms of order $|\mbb_{aIi}|^2/(\m_I^2+|m_n|^2)$.
\Eq{flavor} is the starting point for studying the phenomenology of
our model. Due to our convention for the flavor basis of the
left--handed neutrinos (leading to diagonal charged lepton Yukawa
couplings), the $\nu_i$ are the neutrino ``flavor eigenstates''. They
are expressed as a superposition of three light Majorana mass
eigenstates ($\nu_{0i}$) and a series of left--handed components of
heavy Dirac mass eigenstates ($\nu_{aIn}$). In the limit
$|\mbb_{aIi}|^2/(\m_I^2+|m_n|^2)\ll 1$, the ``heavy'' contribution to
the states $\nu_i$ is small, so that the flavor eigenstates are mainly
light. Therefore, on one hand the model can potentially accommodate the
standard oscillation phenomenology, which mainly involves the three
light Majorana eigenstates $\nu_{0i}$ and the $3\times 3$ mixing
matrix $U$. On the other hand, as we will see, the small admixture of
the Kaluza--Klein states account for the KARMEN anomaly.

\medskip

Let us discuss the results that we have obtained so far from the
general perspective of brane--world models for neutrino masses.  We
have found three exactly massless Weyl fermions as given by
eq.~\eqref{massless}. In general they are superpositions of the
electroweak eigenstates $\n_i$ on the brane and the Kaluza--Klein
states ${\eta'_{aIn}}$.  The relative weight of these two
contributions is controlled by the Dirac masses $\m_I$. For vanishing
Dirac masses $\m_I\rightarrow 0$ the right hand side of
eq.~\eqref{massless} is dominated by the terms corresponding to the
Kaluza--Klein zero mode ${\eta'_{aI0}}$. Hence, in this limit, the
appearance of massless modes is not surprising. The massless
states~\eqref{massless} are simply linear combinations of the bulk
zero modes in this case. The situation becomes less trivial once we
switch on $\m_I$. Then, there are no massless bulk modes any more
since, from eq.~\eqref{kkmass}, the Kaluza--Klein spectrum is bounded
from below by $\m_I$.  To summarize, despite the introduction of a
bulk Dirac mass term, the symmetries of the higher--dimensional theory
(Lorenz invariance and lepton number) lead to massless eigenstates
which are non--trivial linear combinations of the brane and the bulk
fields. We believe this can be of considerable relevance for
brane--world models of neutrino masses, in general.  A final point
concerns the order of magnitude of the bulk Dirac mass $\m$. As
stands, the natural value of this mass is probably the string scale
$M_{\rm s}$. Of course, in this case, the bulk neutrinos would be too
heavy to be relevant for any neutrino physics. If the string scale is
of order $\TeV$ some bulk states might receive smaller masses of
order, say, $\MeV$ due to small couplings. In addition, if the bulk
fermions are interpreted as superpartners of string moduli they
receive masses only after supersymmetry breaking, which may also
account for their smallness. 

\medskip

So far, we have simply assumed that the massless states acquire a
small mass through a breaking of the U(1) symmetry.  However, we have
not explicitly shown how light neutrino masses and mixings needed for
the standard neutrino phenomenology can be generated. While this is
certainly appropriate for our main purpose it is, of course, important
to show how small but non--vanishing neutrino masses can be
incorporated into the model. This is what we will do in the remainder
of the section.

In order to generate neutrino masses we need to break the lepton
number symmetry that we have imposed on our model. Rather than
presenting a complete model of how this can be realized spontaneously,
we simply parameterize this breaking by adding the following Majorana
mass terms to
the bulk action~\eqref{Sbulk}
\begin{equation}
 S_{\rm bulk,L} =
                  \int d^4xd^2y
                  \left[-\frac{1}{2}M_{IJ}^{(A)}\bar{\Psi}_I^c 
                  \Psi_J -\frac{1}{2}M_{IJ}^{(S)}\bar{\Psi}_I^c\G_7\Psi_J+
                  \mbox{h.c.}\right]\; ,
\end{equation}
with $M_{IJ}^{(A)}$ antisymmetric and $M_{IJ}^{(S)}$ symmetric. 
They violate lepton number by two units. Furthermore, we have to amend
the brane action~\eqref{Sbrane} by
\begin{equation}
 S_{\rm brane,L} = \int_{\{y=0\}}d^4x\left[-\frac{l_{aIi}}{M_{\rm s}}
                   \bar{\Psi}^c_{aI}L_iH+\mbox{h.c.}\right]
\end{equation}
which violates lepton number by two units as well.
Recall that $\Psi_{aI}$, where $a=+,-$ are the two
four--dimensional Dirac spinors contained in the six--dimensional
bulk spinors $\Psi_I$. As before, after electroweak symmetry breaking,
we introduce the mass
parameters
\begin{equation}
 \tilde{m}_{aIi} = \frac{l_{aIi}v}{2\p RM_{\rm s}}\; .
\end{equation}
Using the expansion~\eqref{exp} of the bulk fermions we then find the
additional four--dimensional terms 
\bea 
\cL_{\rm bulk,L} &=&
\sum_{n\in\Z^2}\left[ M_{IJ}\eta_{+In}\eta_{-J-n}-
M^\dagger_{IJ}\x_{+In}\x_{-J-n}+\mbox{h.c.}\right] \label{LbuL}\\
\cL_{\rm brane,L} &=& \sum_{n\in\Z^2}\left[\tilde{m}_{+Ii}\x_{+In}
+\tilde{m}_{-Ii}\x_{-In}\right]\n_i
+\mbox{h.c.}
\label{LbrL} 
\eea 
where the Majorana mass matrix
$M$ is defined by $M=M^{(S)}+M^{(A)}$.  Now our total
four--dimensional effective action consists of the original lepton
number preserving parts~\eqref{Lbu}, \eqref{Lbr} and the above lepton
number violating parts~\eqref{LbuL}, \eqref{LbrL}.  It would clearly
be interesting to thoroughly investigate the neutrino phenomenology
resulting from this action that includes all obvious bulk mass terms
allowed by higher--dimensional Lorenz invariance. We hope to return to
this problem in a future publication. For the purpose of the present
paper, we would merely like to check the order of magnitude of masses
that are induced. To do this, we focus on the case of one flavor in
the bulk as well as on the brane. Assuming that $\mbb_a/\mu\ll 1$ and
$\tilde{m}_a/\mu\ll 1$ we can apply the see--saw formula to find the
light neutrino masses. After summing over all Kaluza--Klein states
with cut--off $M_{\rm s}$ we find for the mass $m_0$ of the previously
massless states $\n_{0i}$ 
\bea 
m_0 &=&
(-\tilde{m}_-\mbb_-\mu -\tilde{m}_-\tilde{m}_+M
+\mbb_-\mbb_+M^*-\tilde{m}_+\mbb_+\mu ) \nonumber \\ 
&&\quad\times\left(\frac{2}{\mu^2+|M|^2}+2\p
R^2\ln\frac{M_{\rm s}^2}{\mu^2+|M|^2+1/R^2}\right)\; .
\label{m0} 
\eea 
As expected, this vanishes if we restore lepton number by setting
$\tilde{m}_a=M=0$. The term in the first parenthesis together with the
first term in the second parenthesis correspond to what one normally
expects for a see--saw suppressed neutrino mass. In fact, these terms
describe the contribution from the lightest Kaluza--Klein state with
mode number $n=0$. However, in addition we have a large number of
higher Kaluza--Klein states each of which contributes to the light
neutrino mass. The net effect of all these states is encoded in the
last term in eq.~\eqref{m0}. This term is proportional to $R^2$ rather
than the explicit mass scales $\mu$ or $M$. Hence it corresponds to a
suppression by the mass scale $1/R$ associated to the size of the
additional dimensions. As we will see, in our context, this
suppression is sufficient to obtain reasonable small neutrino masses
under plausible assumptions. Eq.~\eqref{m0} also illustrates the above
mentioned decoupling of neutrino masses and neutrino--bulk mixing. The
former depend in particular on the lepton number violating quantities
$\tilde{m}_\pm$ and $M$. As long as these quantities are sufficiently
small the light neutrinos are still approximately given by
eq.~\eqref{massless} and, hence, they depend on lepton number
conserving parameters only.

\section{Quantitative analysis of the model}
\label{sec:last}

\noindent
We would now like to demonstrate more quantitatively that the
six--dimensional model presented in the previous section can explain
the KARMEN anomaly and is, at the same time, compatible with other
phenomenological constraints.

\medskip

\begin{figure}
\begin{center}
\epsfig{file=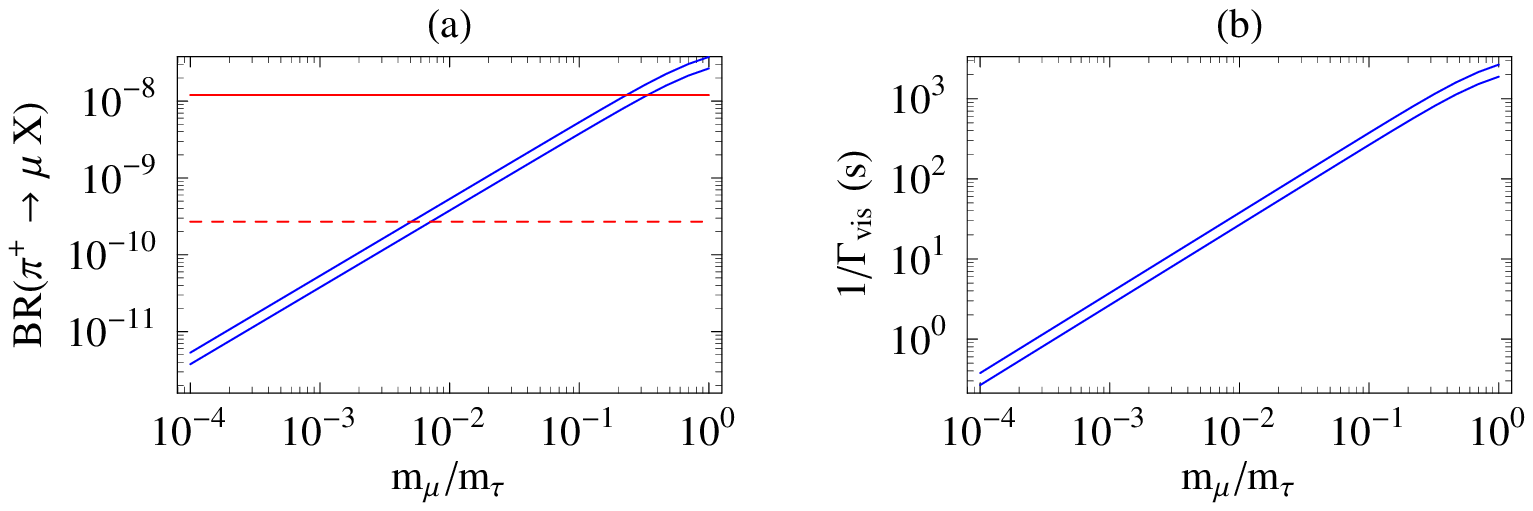,width=\textwidth}
\end{center}
\mycaption{Branching ratio $\text{BR}(\decay)$, (a), and inverse
visible width $\Gamma_{\text{vis}}$, (b), as functions of the ratio
$r=\mbb_\mu/\mbb_\tau$.} 
\label{fig:BRG} 
\end{figure}

Let us start by explaining the requirements on our model that follow from the
KARMEN anomaly. This amounts to specifying the details of the fit that
we have presented in Section~\ref{sec:karmen} and work out some of its
consequences. Let us start with the constraint~\eqref{constr} on the
branching ratio and on the decay width of the $X$ particle that our
model has to satisfy. We should express this constraint in terms
of the parameters of our model. The $X$ particle corresponds to a
mode $n_X$ such that
\begin{equation}
\label{mX}
m_X = \sqrt{\mu^2+|m_{n_{\rm X}}|^2}.
\end{equation}
As we saw in the previous Section, two almost degenerate Dirac mass
eigenstates with opposite internal chirality are associated to this
mode $n_X$, namely $(\tilde\nu^c_{+n_X},\nu_{+n_X})^T$ and
$(\tilde\nu^c_{-n_X},\nu_{-n_X})^T$. The mixing $U^a_{iX}$, where
$a=+,-$ between the SM flavor eigenstates $\nu_i$, $i=e,\mu,\tau$,
and these two states can be read off from \eq{flavor} as
\begin{equation}
\label{mixing}
|U^a_{iX}| = \frac{|\mbb_{ai}|}{\sqrt{\displaystyle\mu^2+|m_{n_X}|^2}}
= \frac{|\mbb_{ai}|}{m_X}.
\end{equation}
However, the two states are not completely degenerate but split at
higher order in the see--saw approximation. For some of the parameter
space this splitting will be smaller than the $X$ particle mass gap
$\dmg\simeq 6\keV$. In this case, both states contribute to the KARMEN
anomaly and can effectively be accounted for by a single state $X$
with mass parameter
\begin{equation}
\label{mdef}
\mbb^2_i = |\mbb_{+i}|^2+|\mbb_{-i}|^2.
\end{equation}
If, on the other hand, the splitting is larger than $\dmg\simeq 6\keV$
one mass eigenstate will be above the threshold for production in the
KARMEN experiment and the anomaly is due to the other state. We will
call the corresponding mass parameter $\mbb$ as well, but it is now
a function of $\mbb_{\pm i}$ generally different from the one given
above. For both cases, we define the mixing matrix elements by
\begin{equation}
\label{mixingsingle}
|U_{iX}| = \frac{\mbb_i}{m_X}\; .
\end{equation}
Through those mixings, the mass parameters $\mbb_i$ control
the $X$ particle production and decay and therefore the number of
anomalous events in the KARMEN experiment. The branching ratio for the decay
$\decay$ is given by
\begin{equation}
\label{BR}
\text{BR}(\decay) = |U_{\mu X}|^2
\frac{\lambda(m^2_\pi,m^2_\mu,m^2_X)}{\lambda(m^2_\pi,m^2_\mu,0)}
\simeq 0.028\, |U_{\mu X}|^2,
\end{equation}
where 
\begin{equation}
\label{phsp}
\lambda(a,b,c)\equiv\left[a(b+c)-(b-c)^2\right] \sqrt{a^2+b^2+c^2-2a
b-2a c-2b c}.
\end{equation}
Possible contributions from bulk dynamics to the visible decay width
are highly suppressed compared to those from standard $W$ and $Z$
exchange. The impact of bulk dynamics on the total width can in
principle be non--negligible but it would not change drastically the
lifetime, which means that it would not affect our conclusions
at all. Neglecting the sub--dominant $X\rightarrow\gamma\nu$ decay, the
visible decay width is given by
\begin{multline}
\label{width}
\Gamma_{\text{vis}} \simeq \Gamma(X\rightarrow e^+e^-\nu) \\ =
\fracwithdelims{(}{)}{m_X}{m_\mu}^5 \left[ |U_{eX}|^2
\left(\frac{1}{4}+s^2_W+2s^4_W\right) + (|U_{\mu X}|^2+|U_{\tau X}|^2)
\left(\frac{1}{4}-s^2_W+2s^4_W\right) \right] \tau_\mu^{-1} \\
\simeq 890 \,\mbox{sec}^{-1} |U_{eX}|^2 + 195\,\mbox{sec}^{-1}
 (|U_{\mu X}|^2+|U_{\tau X}|^2).
\end{multline}
Putting these results together, the constraint~(\ref{constr})
can be expressed in terms of the mixings $|U_{iX}|$ as follows
\begin{multline}
\label{constrU}
(1\dash 2)\,10^{-11} = \text{BR}(\decay)  \Gamma_{\text{vis}}\,
 \mbox{sec} \\
\simeq 25 |U_{\mu X}|^2 |U_{eX}|^2 + 5.5 |U_{\mu X}|^2
(|U_{\mu X}|^2+|U_{\tau X}|^2).
\end{multline}
If we make the plausible assumption
$\mbb_e^2\ll\mbb^2_\mu,\mbb^2_\tau$, it follows that the first term
on the right hand side of the previous equation is sub--dominant.
\Eq{constrU} then represents a constraint on $|U_{\mu X}|$ and
$|U_{\tau X}|$, or, equivalently on $\mbb_\mu$ and $\mbb_\tau$ which reads
\begin{equation}
\label{constrm}
\mbb^2_\mu (\mbb^2_\mu+\mbb^2_\tau) = \left[
(40\dash 47)\keV\right]^4. 
\end{equation}
This constraint relates $\mbb_\mu$
and $\mbb_\tau$ and allows us to compute $\text{BR}(\decay)$ and
$\Gamma_{\text{vis}}$ as a function of the ratio
\begin{equation}
 r\equiv \mbb_\mu/\mbb_\tau\; .
\end{equation}
The result is shown in \Fig{BRG} for the range
$10^{-4}<\mbb_\mu/\mbb_\tau < 1$ which corresponds to the plausible
situation $\mbb_\mu < \mbb_\tau$. The width of the band in these
figures corresponds to the uncertainty on the right hand side of
eq.~\eqref{constrm}. If the ``atmospheric'' mixing angle measured by
SuperKamiokande originated in the diagonalization of the charged
lepton mass matrix, we would expect $\mbb_\mu/\mbb_\tau =
\ord{1}$. This is because the brane--bulk Yukawa coupling was written
in a basis where the charged lepton mass matrix is diagonal. However,
$\mbb_\mu/\mbb_\tau = \ord{1}$ gives a too high branching ratio
$\text{BR}(\decay)$ according to the present PSI upper limit
(solid horizontal line in \Fig{BRG}a). The new expected limit is also
shown (dashed horizontal line in \Fig{BRG}a). The latter would imply
$\mbb_\mu/\mbb_\tau\lesssim 10^{-2}$. Therefore, we focus on the case
where $\mbb_\mu/\mbb_\tau\ll 1$.
Furthermore, a lower constraint on the $\mbb_\mu/\mbb_\tau$ range is
given by the requirement that $\mbb_\tau$ is not too large. If
$\mbb_\tau$ were of the order of the Dirac mass $\mu$ or more,
the $\tau$ neutrino would predominantly consist of a heavy mass
eigenstate, which would be phenomenologically problematic.
Consequently, $r$ should be preferably in the range
\begin{equation}
 r = 10^{-4}\;\mbox{to}\; 10^{-2}\; .\label{rrange}
\end{equation}
Then, the mass parameters $\mbb_i$ of our model as well as the branching
ratio and the decay width of the $X$ particle can be expressed in
terms of $r$ as
\globallabel{numbers}
\begin{gather}
\mbb_\mu = (4.0\dash 4.7)\keV
\fracwithdelims{(}{)}{r}{10^{-2}}^{1/2} \mytag \\
\mbb_\tau = (0.4\dash 0.47)\MeV
\fracwithdelims{(}{)}{10^{-2}}{r}^{1/2} \mytag \\
\text{BR}(\decay) = (3.8\dash 5.3)\,10^{-10}\,
\fracwithdelims{(}{)}{r}{10^{-2}} \mytag \\
\tau_X < \Gamma_{\text{vis}}^{-1} = (27\dash 38)\,\text{sec}\,
\fracwithdelims{(}{)}{r}{10^{-2}}. \mytag
\end{gather}
It remains to fix the radius $R$ of the two additional dimensions.  Of
course, $1/R$ should be smaller than the mass $m_{\rm X}\simeq
33.9\MeV$ of the $X$ particle. Furthermore, as we have seen in
Section~\ref{sec:karmen}, a value of $1/R\simeq 1.6\MeV$ leads to a
complete solution of the fine--tuning problem. However, as explained
there, we need to choose a somewhat larger value $1/R\gtrsim 15\MeV$
to avoid significant contributions from modes lighter than the $X$
particle to the KARMEN spectrum. In summary, we therefore require
\begin{equation}
 15\MeV\lesssim 1/R < 33.9\MeV\; .
\end{equation}
The fine--tuning is smallest at the lower end of this range which is
why we have chosen $1/R\simeq 15\MeV$ in the fit performed in Section
\ref{sec:karmen}. Finally, we have to fix the order of magnitude of our
Dirac mass $\m$. To avoid too large admixture of the lightest
Kaluza--Klein mode we have to require that $\m\gg\mbb_\t$. In
addition, of course, $\m$ has to be smaller than the $X$ particle mass.
Together, this leads to 
\begin{equation}
 \mbb_\t\ll\m <33.9\MeV\; .
\end{equation}
We remark that this allows for the plausible
value $\m = O(1/R)$ for the Dirac mass. Finally, we should check the
assumption made in the previous section about the unitarity of the
matrix $N$ and the smallness of the heavy states contribution to
the SM neutrinos. Both issues are related to the square of the
normalization matrix $N^\dagger N$ defined in eq.~\eqref{N} which
should be close to unity. For our favorite values
$1/R\simeq\m\simeq 15\MeV$ and $M_{\rm s}=1\TeV$ this square is given
by
\begin{equation}
 N^\dagger N\simeq 1+0.05\left(\frac{10^{-2}}{r}\right)\; 
\end{equation}
It constraints $r$ to be in a subset of the range~\eqref{rrange} which,
however, is still comfortably large.

\medskip

Now that we have basically fixed all the lepton number conserving
quantities in our model we should analyze what we need to do to
generate small masses for the neutrinos. In the previous section we
have shown that we can generate neutrino masses by introducing a lepton
number breaking bulk Majorana mass $M$ and lepton number violating
brane--bulk masses $\tilde{m}_\pm$. Both mass terms violate lepton
number by two units. For the simple case of one neutrino flavor the
resulting neutrino mass has been given in eq.~\eqref{m0}. Using our
favorite values $1/R\simeq\m\simeq 15\MeV$ and $\mbb_\pm\simeq\mbb_\t$
where $\mbb_\t$ has been specified in eq.~\eqref{numbers} we find
\begin{equation}
 m_0\simeq\left(\frac{10^{-2}}{r}\right)\left(
 -2\frac{\tilde{m}_\t}{\mbb_\t}+\frac{M}{\m}\right)(1.5\dash 2.0)\MeV\; .
\end{equation}
Note that the two ratios $\tilde{m}_\t /\mbb_\t$ and $M/\m$ of lepton
number violating and conserving quantities on the brane and in the
bulk, respectively, enter with the same strength. This fits nicely
to the fact that they both break lepton number by two units. More
quantitatively, to have a neutrino mass $m_0\lesssim 1\eV$ we need both
ratios to be less than roughly $10^{-4}r$. Such a number may, for
example, arise from a suppression of the form $<\f >/M_{\rm s}$ where
$\f$ is a boson that carries lepton number and takes a VEV of the
order, say $1/R$.

\medskip

Although the main focus of the paper is on the particle physics
aspects of our model, we would like to briefly address constraints
from big--bang nucleosynthesis and astrophysics. 
In the early universe, the Kaluza--Klein mode with mode number
$n$ and mixing $U_{i n}$ decouples at a temperature of
\begin{equation}
 T_{\rm dec,n} \simeq 1\MeV |U_{i n}|^{-2/3}\; .
\end{equation}
From eq.~\eqref{numbers} the largest mixing angles are those for
$i=\t$. Using these angles and $\mu=1/R\simeq 15\MeV$ as above we find
\begin{equation}
 T_{\rm dec,n} \simeq
 (10\text{--}11)\MeV\left(\frac{r}{10^{-2}}\right)^{1/3}
 (1+n^2)^{1/3}\; .
\end{equation}
It follows than the decoupling temperature is lower than the mass for
all Kaluza--Klein modes. In particular, the decoupling temperature
increases with a smaller power of the mode number $n$ than the masses
$M_n\simeq 15\MeV\sqrt{1+n^2}$ of the Kaluza--Klein modes.  As a
consequence, modes with large $n$ decouple when they are highly
non--relativistic and are strongly diluted. Unfortunately, this
suppression is not strong enough for low $n$ modes. Therefore, we have
to assume that the bulk is empty at some temperature $T<T_{\rm
dec,0}\simeq 10\MeV$ as it is customary for models with large
additional dimensions~\cite{Arkani-Hamed:98b}. Taking this temperature
sufficiently low (but above $1\MeV$, of course) recreation of
Kaluza--Klein particles is Boltzmann--suppressed. As a consequence,
their thermal production rate can always be kept below the Hubble
rate.

\medskip

Supernova cooling provides another constraint on our model. For the
single--particle explanation of the KARMEN anomaly the supernova
energy loss induced by the $X$ particle has been estimated in
ref.~\cite{DHRS}. From this estimate, it is concluded there, that the
single particle explanation of the KARMEN anomaly is ruled out.  Our
model is even more problematic since we have a tower of particles
rather than just a single one. However, a closer analysis of the
supernova energy loss~\cite{RS} is likely to weaken the bound given in
ref.~\cite{DHRS}.  This together with the assumption of a moderately
lower supernova temperature might render our model consistent with the
supernova constraint.  In any case, we believe that our model is of
interest from the viewpoint of particle physics and should be compared
with the direct experimental information on neutrino properties.

\section{Conclusions}

In this paper, we have shown that the KARMEN anomaly can be understood
in the context of a six--dimensional brane--world model. The slow
particle $X$, responsible for the anomaly, is identified with a
specific Kaluza--Klein excitation of a bulk fermion that, from a
four--dimensional point of view, appears as a sterile neutrino.  We
have pointed out that in any interpretation of the KARMEN anomaly
based on a single slow particle produced in the $\decay$ decay, the
$X$ mass is fine--tuned to the mass difference $m_\pi-m_\mu$ with an
accuracy of order $10^{-4}$. This means that $\dmg/(m_\pi-m_\mu)\sim
1.8\, 10^{-4}$, where $\dmg=(m_\pi-m_\mu) - m_X$. Such a problem can
be significantly alleviated, although only partially solved, in the
approach we propose, since the $X$ particle is just one of the many
excitation of the bulk fermion. It is therefore relatively more
probable that one out of the many states has a mass that falls into
the critical range.

The phenomenology of the model with respect to the KARMEN experiment
depends on the average separation $\dms$ between two Kaluza--Klein
states at the scale $m_X$. In the limit where $\dms$ is large, that is
$\dms=\ord{m_\pi-m_\mu}$, the model is indistinguishable from models
with a single sterile neutrino. In this limit, although the
fine--tuning is not improved, we still have a model that provides a
theoretically well--motivated origin for the $X$ particle. Moreover,
as a quite non--trivial feature, the model allows to incorporate the
large mass scale of the $X$ particle as well as the small neutrino
masses needed for the standard oscillation phenomenology. The limit of
small $\dms$, that is $\dms=\ord{\dmg}$, is ruled out because it would
lead to a modification of the KARMEN time spectrum in the region
$(0.6\dash 3.1)\us$. For intermediate values of $\dms$,
$\dms=\ord{1}\MeV$ or more, the fine--tuning problem is significantly
alleviated and the particles lighter than $X$ could give a small
contribution to the early KARMEN time spectrum. Therefore, a detailed
study of the time distribution of events at early times after the end
of the proton pulse could test the model if the mass separation $\dms$
happens not to be too large. An upgraded detector with tracking
capability devoted to the study of the anomaly could certainly explore
this possibility.  Needless to say, the detection of a signal at early
times would represent a strong hint for the existence of a
brane--world.

On the model building side, we have identified two possible
structures for the sizes of the additional dimensions. The first case
represents an effectively five--dimensional model with intermediate
string scale and the remaining five dimensions being close to the
string scale. The second one has a $\TeV$ string scale, two large
dimensions in the $\MeV^{-1}$ range and the scale of the remaining
four dimensions being two to three orders of magnitude
larger. Therefore, in this case the space--time is effectively
six--dimensional in an intermediate energy range. We have described in
detail a six--dimensional example. As in all brane--world neutrino
models, we have considered bulk neutrinos whose Lorenz invariant
Dirac and Majorana masses are suppressed relative to the string
scale. In particular, we have taken the magnitude of the Dirac mass
term to be of the order of the inverse size of the additional
dimensions and we have further suppressed the Majorana mass term by
means of a U(1) symmetry (lepton number). Despite the introduction of
the bulk Dirac mass term, we have found that the model accommodates
three light Majorana neutrinos which would be massless in the limit of
unbroken U(1) symmetry. The light states are predominantly given by
the electroweak eigenstates with small but sizeable admixture of
Kaluza--Klein modes. We have therefore attributed the solar and
atmospheric neutrino phenomenology to those light states. On the other
hand the small heavy component of the flavor eigenstates accounts for
the KARMEN anomaly. Therefore, the introduction of the Dirac mass term
makes our model rather different from the models considered in the
literature so far, where the massless modes either do not exist or
decouple from the left--handed neutrinos.  Moreover, the neutrino
masses and the mixing of neutrinos and bulk states are controlled by
different parameters. We believe that such a scheme can be of general
interest for brane--world models of neutrino masses.

\section{Acknowledgments}
We thank Pierre Ramond, Graham Ross and Subir Sarkar for useful
discussions and suggestions. This work is supported by the TMR Network
under the EEC Contract No.~ERBFMRX--CT960090.

\providecommand{\href}[2]{#2}\begingroup\raggedright\endgroup

\end{document}